\documentclass[thmsa,final,11pt]{article}
\begin{document}

\bigskip

%TCIMACRO{
%\TeXButton{TeX field}{
%
%
%\bigskip %
%}}%
%BeginExpansion

\bigskip %
%
%EndExpansion

\begin{center}
{\LARGE \textbf{Partial algebraization and a q-deformed harmonic oscillator}}
\end{center}

%TCIMACRO{
%\TeXButton{TeX field}{\bigskip %
%}}%
%BeginExpansion
\bigskip %
%
%EndExpansion

\begin{center}
{\Large Abilio De Freitas}$\footnote{%
Present address: UCLA,\ Physics Deparment, Box 951361. Los Angeles,
CA-90095-1361.}${\large \ and }{\Large Sebasti\'{a}n Salam\'{o}}$\footnote{%
E-mail adress: ssalamo@ fis.usb.ve}$

\vspace{0.5cm} Universidad Sim\'{o}n Bol\'{i}var, Departamento de F\'{i}sica,%
\\[0pt]
Apartado Postal 89000 ,Caracas,Venezuela
\end{center}

\medskip
%TCIMACRO{
%\TeXButton{TeX field}{\bigskip %
%}}%
%BeginExpansion
\bigskip %
%
%EndExpansion

\bigskip

\begin{center}
{\Large \textbf{Abstract}}
\end{center}

From the algebraic treatment of the quasi-solvable systems, and a
q-deformation of the associated $su(2)$ algebra, we obtain exact solutions
for the q-deformed Schr\"{o}dinger equation with a 3-dimensional q-deformed
harmonic oscillator potential.

%TCIMACRO{
%\TeXButton{TeX field}{\bigskip   %
%}}%
%BeginExpansion
\bigskip   %
%
%EndExpansion

%{\Large Introduction}

\section{Introduction}

\smallskip In spite of the fact that the Schr\"{o}dinger equation was
introduced many years ago, the development of techniques, and specially,
algebraic techniques, elaborated with the purpose of solving this equation,
is still a living subject. Between the most remarkable and general works
done to this date we can mention the following ones: 1. - The SO(2,2)
treatment for the hypergeometric Natanzon potentials [1], studied by
Alhassid et al [2]. 2. - The SUSYQM method for the shape-invariant
potentials by Gendeshtein, Cooper, Dutt [3]. 3. - The SO(2,1) treatment for
the hypergeometric-confluent Natanzon potentials studied in [4]. 4. - The
partial algebraization method, which allows to obtain part of the spectrum
for the so-called quasi-solvable systems, as we can see in [5]. We should
also mention the works done by Kosteleck\'{y}, Nieto and Man'ko, about
algebraic descriptions including spin [6].

The following question arises: what can be done in the context of
deformations? In other words, Is it possible to extend some of the previous
approaches to include quantum groups?

At this point we mention the works done by: Dayi and Duru (Coulomb and Morse
potentials) [7], Cooper (Morse) [8], Daskaloyannis et al (Morse) [9], De
Freitas and Salam\'{o} (P\"{o}schl-Teller I and II) [10]. The references
[7], [8] and [9] are related to the algebraic treatments of the
``classical'' systems.

In particular, it would be interesting to obtain algebraic solutions for the
q-Schr\"{o}dinger equation. This is, a Schr\"{o}dinger equation in which the
usual commutation relation between the position and momentum operators, $%
\left[ X,P\right] =i$, is substituted by $XP-qPX=i$, which can be done by
associating the momentum operator to the deformed derivative, instead to the
usual one, as we will see later on.

In this paper we work out such an algebraic method for the particular case
in which the potential is a 3-dimensional q-harmonic oscillator. By
q-harmonic oscillator we mean a q-dependent potential that reduces to the
usual harmonic oscillator in the proper limit of $q$ ($q\rightarrow 1$). The
method is strongly based on the ideas of partial algebraization. It relies
on the use of a particular deformation of the $su(2)$ algebra (a quadratic
one), which we call $\mathcal{D}_{q}^{n}(su(2))$. In the same spirit of the
partial algebraization method, we express the Hamiltonian of the system in
terms of the generators defining this $\mathcal{D}_{q}^{n}(su(2))$
q-algebra. This allows us to obtain exact solutions for part of the spectrum
as well as the corresponding eigenfunctions.

This paper is organized as follows: In section 2 we make a review of the
partial algebraization method. This is required to understand what follows
in the paper, though it is a somewhat long section, so in order to avoid
confusion, the reader should keep in mind that our goal is the
algebraization of the q-Schr\"{o}dinger equation with a q-harmonic
oscillator. This is the central point of section 3 as well as the main part
of this paper. In the mentioned section 3 we define and construct the $%
\mathcal{D}_{q}^{n}(su(2))$ algebra. After rotating the q-Schr\"{o}dinger
equation, in analogy with what is done in the partial algebraization method,
we continue the analogy, relating this rotated equation with a quadratic
plus a linear expression of the generators defining $\mathcal{D}%
_{q}^{n}(su(2))$. In section 4 we use a particular representation of $%
\mathcal{D}_{q}^{n}(su(2))$ to show how we can get the so desired exact
solutions of our q-Schr\"{o}dinger equation. Next, some particular examples
are shown. Finally, in section 5 we discuss the obtained results.

\bigskip

%\smallskip {\Large Partial Algebraization}

\section{Partial Algebraization}

\smallskip

The idea of the method of partial algebraization [5] consists in to express
the Schr\"{o}dinger equation
\begin{equation}
\left( -\frac{1}{2}\frac{d^{2}}{dx^{2}}+(V(x)-E)\right) \Psi (x)=0
\end{equation}
in terms of the generators of SU(2). This can be done by first doing a
non-unitary transformation on (1) in the following way

\begin{equation}
\Psi (x)=\widetilde{\Psi }(x)\exp (-a(x))
\end{equation}
thus (1) is now given by

\begin{equation}
H_{G}\widetilde{\Psi }(x)=E\widetilde{\Psi }(x)
\end{equation}
with

\begin{equation}
H_{G}=-\frac{1}{2}\frac{d^{2}}{dx^{2}}+A(x)\frac{d}{dx}+\Delta V
\end{equation}
where

\begin{equation}
\Delta V=V(x)+\frac{1}{2}\frac{dA(x)}{dx}-\frac{1}{2}A^{2}(x)
\end{equation}
and\ $A(x)$ is defined as

\begin{equation}
A(x)=\frac{da(x)}{dx}
\end{equation}
The realization for the SU(2) algebra is taken to be

\begin{equation}
J_{+}=2ju-u^{2}\frac{d}{du},\quad J_{0}=-j+\frac{d}{du},\quad J_{-}=\frac{d}{%
du}
\end{equation}
with the usual commutation relations: $[J_{+},J_{-}]=2J_{0},\
[J_{0},J_{+}]=J_{+},\ [J_{0},J_{-}]=-J_{-}.$ The index $\ j$ \ labels the
representations of dimensions $2j+1.$ The carrier space is given by
monomials of degree $n$ on the variable $u$, i.e., $u^{n}.\ $By other hand,
since the eigenvalues of\ $J_{0}$ lies between $-j$ and\ $\ j$ , then $0\leq
n\leq 2j$, thus the representation space $R^{j}$ is given by

\begin{equation}
R^{j}=\{1,u,u^{2},...u^{2j}\}
\end{equation}

The algebraization of the Schr\"{o}dinger equation is made assuming that the
Hamiltonian $H_{G}$ given in (4), may be written as

\begin{equation}
H_{G}=\sum_{a,b=0,\pm ,a\geq b}C_{ab}J_{a}J_{b}+\sum_{a,b=0,\pm }C_{a}J_{a}
\end{equation}
where $C_{ab}$ and $C_{a}$ are constants. We obtain

\begin{equation}
H_{G}=-\frac{1}{2}P_{4}(u)\frac{d^{2}}{du^{2}}+P_{3}(u)\frac{d}{du}+P_{2}(u)
\end{equation}
with $P_{4}(u)$,$\ P_{3}(u)\ $and $P_{2}(u)$ given by

\begin{eqnarray}
P_{4}(u) &=&-2C_{++}u^{4}+2C_{+-}u^{3}+2(C_{+-}-C_{00})u^{2}-2C_{0-}u-2C_{--}
\nonumber \\
P_{3}(u) &=&2C_{++}(1-2j)u^{3}+((3j-1)C_{+0}-C_{+})u^{2} \\
&&+(2jC_{+-}+(1-2j)C_{00}+C_{0})u-jC_{0-}+C_{-}  \nonumber \\
P_{2}(u) &=&2j(2j-1)C_{++}u^{2}+2j(C_{+}-jC_{+0})u+j^{2}C_{00}-jC_{0}
\nonumber
\end{eqnarray}
A more compact form can be achieved if we define a new set of parameters as
follows

\begin{eqnarray}
a_{4} &=&-2C_{++},\quad a_{3}=2C_{+0},\quad a_{2}=2(C_{+-}-C_{00}),\quad
a_{1}=-2C_{0-}  \nonumber \\
a_{0} &=&-2C_{--},\quad b_{2}=C_{+}-jC_{+0},\quad
b1=2jC_{+-}+(1-2j)C_{00}+C_{0}  \nonumber \\
b_{0} &=&-jC_{0-}+C_{-},\quad c_{0}=j^{2}C_{00}-jC_{0}
\end{eqnarray}
\newline
therefore (11) is now given as

\begin{eqnarray}
P_{4}(u) &=&a_{4}u^{4}+a_{3}u^{3}+a_{2}u^{2}+a_{1}u+a_{0}  \nonumber \\
P_{3}(u) &=&(2j-1)a_{4}u^{3}+((j-\frac{1}{2})a_{3}-b_{2})u^{2}+b_{1}u+b_{0}
\\
P_{2}(u) &=&j(1-2j)a_{4}u^{2}+2jb_{2}u+c_{0}  \nonumber
\end{eqnarray}

We now proceed to identify the expressions for $H_{G}$ given in (4) and
(10). Since the variables $x$ and $u$ are involved, a change of variables
must be done, as follows

\begin{equation}
x=\int P_{4}^{-\frac{1}{2}}(u)du=\phi (u)
\end{equation}
which also allows to eliminate the term multiplying the second derivative in
(10). In this way we get

\begin{equation}
H_{G}=-\frac{1}{2}\frac{d^{2}}{dx^{2}}+\frac{(4P_{3}(u)+P_{4}^{^{\prime
}}(u))}{4P_{4}^{\frac{1}{2}}(u)}\frac{d}{dx}+P_{2}(u)
\end{equation}
The substitution $\ u=\phi ^{-1}(x)$ has to be done in the polynomials
occurring in (10). Furthermore, $P_{4}^{^{\prime }}(u)$ is the derivative of
$P_{4}(u)$ respect to $u$ with the subsequent substitution of\ $u$ as a
function of $\ x.$

We can see now more clearly the reason for transforming the Schr\"{o}dinger
equation with the non-unitary operator $e^{-a(x)},$ since as a consequence
of that, we get a term with first derivative which makes possible the
identification of (4) and (10). From this identification we get

\begin{equation}
A(x)=\frac{4P_{3}(u)+P_{4}^{^{\prime }}(u)}{4P_{4}^{\frac{1}{2}}(u)}
\end{equation}
and

\begin{equation}
\Delta V=V(x)+\frac{1}{2}A^{^{\prime }}(x)-\frac{1}{2}A^{2}(x)=P_{2}(x)
\end{equation}

Substituting (16) in (17) we get the general expression for the potentials
which may be algebraized by su(2)
\begin{eqnarray}
V(x) &=&P_{2}(u)+\frac{1}{8P_{4}(u)}(P_{4}^{^{\prime }}(u)+4P_{3}(u))^{2}-%
\frac{1}{8}P_{4}^{^{\prime \prime }}(u)-\frac{1}{2}P_{3}^{^{\prime }}(u)
\nonumber \\
&&+(P_{4}^{^{\prime }}(u)+4P_{3}(u))\frac{P_{4}^{^{\prime }}(u)}{16P_{4}(u)}
\end{eqnarray}
The set of parameters characterizing this potential are: $a_{4},\ a_{3},\
a_{2},\ a_{1},\ a_{0},\ b_{2},\ $\newline
$b_{1},\ b_{0}\ $and $\ c_{0}.$

Let us see now how this algebraization lead us to the energy spectrum and
the corresponding eigenfunctions of the potential obtained in (18). Since
the carrier space of $su(2)$ is given by (8), then a basis for the
eigenfunctions of $H_{G}$ is
\begin{equation}
\{\widetilde{\Psi }\}=\{1,u,u^{2},...u^{2j}\}\cup \{\widetilde{\Psi }_{2j+2},%
\widetilde{\Psi }_{2j+3}...\}
\end{equation}
where $\{\widetilde{\Psi }_{2j+2},\widetilde{\Psi }_{2j+3}...\}$ is
orthogonal to $R^{j}.$ So, the matrix representation of $H_{G}$ is given by

\begin{equation}
H_{G}=\left(
\begin{tabular}{cc}
$H_{G1}$ & $0$ \\
$0$ & $H_{G2}$%
\end{tabular}
\right)
\end{equation}
where $H_{G1}$ is the matrix associated to $R^{j}$ and $H_{G2}$ is an
infinite matrix associated to the second subspace, which is unknown.
Therefore, since $H_{G1}$ is a hermitic matrix of dimension $(2j+1)\times
(2j+1)$, it is possible to diagonalize it, obtaining a part of the energy
spectrum and their corresponding eigenfunctions. This is the reason why the
potentials treated in this way are known as quasi-solvables, since only a
part of the spectra may be obtained. Nevertheless, it is possible to find
the complete spectra with this technique for some potentials, as we will see
later.

The eigenfunctions of $H_{G1}$ will be linear combinations of elements of $%
R^{j}$, therefore if$\ $ $H_{G1}\widetilde{\Psi }=E\widetilde{\Psi }$, then $%
\widetilde{\Psi }(u)=P_{2j}(u)$, where $P_{2j}(u)$ is a polynomial on $u$ of
degree $2j$. Using (2,6,16), the eigenfunctions of $H_{G1}$ turns out to be
of the form
\begin{equation}
\Psi (u)=\widetilde{\Psi }(x)e^{-a(x)}=P_{2j}(u(x))e^{-\int \frac{%
(4P_{3}+P_{4}^{^{\prime }})}{4P_{4}^{\frac{1}{2}}}dx}
\end{equation}
and the corresponding eigenvalues are obtained after the diagonalization of $%
H_{G1}.$

Let us see one example. Let us take: $a_{4}=\ a_{3}=\ a_{1}=\ a_{0}=c_{0}=0.
$ Thus $P_{4}(u)=a_{2}u^{2}$, from (14) we get

\[
u=e^{kx}
\]
with $k=\sqrt{a_{2}}.$ The potential for this set of parameters is obtained
from (18) and is given by

\begin{eqnarray}
V(x) &=&\frac{b_{2}^{2}}{2k^{2}}e^{2kx}+\frac{b_{2}}{k^{2}}%
(2jk^{2}-b_{1})e^{kx}+\frac{b_{0}}{k^{2}}(k^{2}+b_{1})e^{-kx}  \nonumber \\
&&+\frac{b_{0}^{2}}{2k^{2}}e^{-2kx}+\frac{1}{8k^{2}}\left(
(k^{2}+2b_{1})^{2}-8b_{0}b_{2}\right)
\end{eqnarray}
>From (16)

\[
A(x)=\frac{1}{2k}(k^{2}+2b_{1})+\frac{b_{0}}{k}e^{-kx}-\frac{b_{2}}{k}e^{kx}
\]
hence, after using (6)

\begin{equation}
a(x)=\frac{1}{2k}(k^{2}+2b1)\ x-\frac{b_{0}}{k^{2}}e^{-kx}-\frac{b_{2}}{k^{2}%
}e^{kx}
\end{equation}
$H_{G}$ is then obtained from (15),

\[
H_{G}=-\frac{1}{2}a_{2}u^{2}\frac{d^{2}}{du^{2}}+(-u^{2}b_{2}+b_{1}u+b_{0})%
\frac{d}{du}+2jb_{2}u
\]
In the case\ \ $j=0$, the representation space will be simply $R^{0}=\{1\}$,
and we will find only one energy eigenvalue: $E_{0}=0$ \ with $\Psi
_{0}=e^{-a(x)}$ its corresponding eigenstate. If\ \ $j=\frac{1}{2}$ we have
then $\ R^{1/2}=\{1,u\}.$ Representing each element of this space (i.e., $1$
and $u$) by $\left| 0\right\rangle $ and $\left| 1\right\rangle $
respectively, we have then the following matrix elements: $\left\langle
0\left| H_{G}\right| 0\right\rangle =0$, $\ \left\langle 0\left|
H_{G}\right| 1\right\rangle \ =b_{0}$ , $\left\langle 1\left| H_{G}\right|
0\right\rangle =b_{2}$, and\ $\ \left\langle 1\left| H_{G}\right|
1\right\rangle =b_{1}.$ So, the expression for $H_{G1}$ in a matrix
representation is given by

\[
H_{G1}=\left(
\begin{tabular}{cc}
$0$ & $b_{0}$ \\
$b_{2}$ & $b_{1}$%
\end{tabular}
\right)
\]
whose eigenvalues are

\[
E_{\pm }=\frac{1}{2}b_{1}\pm \frac{1}{2}\sqrt{b_{1}^{2}+4b_{2}b_{0}}
\]
with the corresponding eigenstates

\[
\Psi _{\pm }(x)=N\ (1+\frac{E_{\pm }}{b_{0}}e^{kx})\ e^{-a(x)}
\]
where $a(x)$ is given in (23) and $N$ is a normalization constant. Notice
that if we set $\ b_{2}=0$ in (22) we obtain the Morse potential which is
exactly solvable. We expect then that with a proper choice of parameters one
could obtain certain potentials that can be solved in a closed form, this is
the point that we would like to discuss now.

We can find a general class of potentials that are exactly solvable through
this method if we take the potential (18) to be independent of $j$. After a
tedious calculation one find the following conditions for the coefficients
occurring in (18)

\begin{equation}
\begin{array}{l}
2(a_{2}b_{2}+a_{4}b_{0})+a_{3}b_{1}-a_{4}a_{1}=0, \\
2(2a_{1}b_{2}+a_{3}b_{0})-a_{3}a_{1}-4a_{4}a_{0}=0 \\
a_{0}(a_{3}-2b_{2})=0,\quad a_{3}b_{2}+2a_{4}b_{1}=0 \\
a_{4}a_{0}=0,\quad a_{4}a1=0,\quad a_{3}^{2}-4a_{4}a_{2}=0
\end{array}
\end{equation}
This system has only one interesting solution and is given by

\begin{equation}
a_{4}=a_{3}=b_{2}=0
\end{equation}
Substituting (25) in (18) we obtain for $V(x)$

\begin{eqnarray}
V(x) &=&\frac{%
(2(a_{2}+2b_{1})u+a_{1}+4b_{0})(2(3a_{2}+2b_{1})u+3a_{1}+4b_{0})}{%
32(a_{2}u^{2}+a_{1}u+a_{0})}  \nonumber \\
&&-\frac{1}{2}b_{1}-\frac{1}{4}a_{2}+c_{0}
\end{eqnarray}
with
\begin{equation}
\frac{du}{dx}=\sqrt{a_{2}u^{2}+a_{1}u+a_{0}}
\end{equation}

The expression for $V(x)$ given in (26) is the most general one that can be
completely algebrized using the su(2) algebra.

Under the conditions that we have imposed , the ``rotated'' Hamiltonian $%
H_{G}$ (15) turns out to be

\begin{equation}
H_{G}=-\frac{1}{2}(a_{2}u^{2}+a_{1}u+a_{0})\frac{d^{2}}{du^{2}}%
+(b_{1}u+b_{0})\frac{d}{du}+c_{0}
\end{equation}

So, if\ $\left| n\right\rangle $ represents the element $u^{n}$ of the
carrier space $R^{j}$, with\ $j$ arbitrary, the matrix elements of\ $H_{G}$
are given by

\begin{eqnarray}
\left\langle m\left| H_{G}\right| n\right\rangle &=&(-\frac{1}{2}%
a_{2}n(n-1)+nb_{1}+c_{0})\delta _{n,m} \\
&&+(-\frac{1}{2}a_{1}n(n-1)+nb_{0})\delta _{n-1,m}-\frac{1}{2}%
a_{0}n(n-1)\delta _{n-2,m}  \nonumber
\end{eqnarray}

Therefore, $H_{G}$ is a triangular superior matrix, which means that their
eigenvalues are those of the diagonal, i.e.,
\begin{equation}
E_{n}=-\frac{1}{2}a_{2}n(n-1)+nb_{1}+c_{0}
\end{equation}
\renewcommand{\arraystretch}{1.5}

In the following table we exhibit the cases that can be solved through this
technique, as well as the corresponding parameters.\newline
\begin{tabular}{|c|c|c|c|c|c|c|}
\hline
$\mbox{Potential}$ & $a_{2}$ & $a_{1}$ & $a_{0}$ & $b_{1}$ & $b_{0}$ & $%
c_{0} $ \\ \hline
$\frac{1}{2}\omega ^{2}x^{2}+\frac{l(l+1)}{2x^{2}}$ & $0$ & $1$ & $0$ & $%
2\omega $ & $\frac{l}{2}-\frac{1}{4}$ & $\omega (\frac{1}{2}-l)$ \\ \hline
$\frac{1}{2}A^{2}+\frac{1}{2}B^{2}e^{-2\alpha x}-B(A+\alpha )e^{-\alpha x}$
& $\alpha ^{2}$ & $0$ & $0$ & $\alpha A$ & $-\alpha B$ & $-\frac{\alpha }{2}%
(A+\frac{\alpha }{4})$ \\ \hline
$\frac{1}{2}A^{2}+\frac{1}{2}(B^{2}-A^{2}-\alpha A)\mbox{sech}(\alpha x)^{2}$
& $\alpha ^{2}$ & $0$ & $1$ & $\alpha A-\frac{1}{2}\alpha ^{2}$ & $B$ & $0$
\\
$+$ $\frac{1}{2}B(2A+\alpha )\mbox{sech}(\alpha x)\tanh (\alpha x)$ &  &  &
&  &  &  \\ \hline
$\frac{1}{2}A^{2}+\frac{1}{2}(B^{2}-A^{2}+\alpha A)\mbox{csch}(\alpha x)^{2}$
& $\alpha ^{2}$ & $0$ & $-1$ & $\alpha A-\frac{1}{2}\alpha ^{2}$ & $-B$ & $0$
\\
$-\frac{1}{2}B(2A+\alpha )\coth (\alpha x)\mbox{csch}(\alpha x)$ &  &  &  &
&  &  \\ \hline
\end{tabular}

Introducing these values for the parameters in (30) we can check that we get
the correct expressions for the energy eigenvalues for each potential, see
for example the table in [11].

\bigskip \bigskip

%{\Large Algebrization of the q-Scrh\"{o}dinger eqn. through D}$%
%_{q}^{n}(su(2))$

\section{Algebraization of the q-Scrh\"{o}dinger equation through $\mathcal{D%
}^n_q (su(2))$}

\smallskip \bigskip \smallskip

We would like to use the ideas of the partial algebraization method
presented in the preceding section to solve the q-deformed Schr\"{o}dinger
equation. A deformation is done after associating the momentum operator $P$
to a deformed derivative. Then the standard commutation relation between the
momentum and the position operator $X$ is replaced by

\[
XP-qPX=i
\]
In this way we assume that a q-deformed Schr\"{o}dinger equation is given by

\begin{equation}
\left( -\frac{1}{2}D_{q}^{2}+(V_{q}(x)-E)\right) \Psi _{q}=0
\end{equation}
where the deformed derivative is given by

\[
D_{q}\Phi (x)=\frac{\Phi (x)-\Phi (qx)}{(1-q)\ x}
\]

In this paper we will obtain exact solutions of (31), for the particular
case in which the potential $V_{q}(x)$ is a q-deformed harmonic oscillator,
using algebraic techniques. For this purpose we will use the $\mathcal{D}%
_{q}^{n}(su(2))$ algebra to be defined later on. A realization of this
algebra is obtained from the $su(2)$ algebra that was used in the preceding
section (7) after a change of variables. As it was seen before, we had to
change variables in order to equate the rotated Hamiltonian with the
combinations of generators given in (9). The change of variables that we are
going to use is a simple one, namely $u=x^{n}$. This is done for simplicity
in order to avoid the complications that arise when one consider the general
case. For this case, we have that (7) becomes

\begin{eqnarray}
J_{+} &=&2jx^{n}-\frac{1}{n}\ x^{n+1}\frac{d}{dx}  \nonumber \\
J_{0} &=&-j+\frac{1}{n}\ x\frac{d}{dx} \\
J_{-} &=&\frac{1}{n}\ x^{1-n}\frac{d}{dx}\quad  \nonumber
\end{eqnarray}

If we make a formal replacement of $\frac{d}{dx}\rightarrow D_{q}$ in (32)
we have then

\begin{eqnarray}
J_{+} &=&2jx^{n}-\frac{1}{n}\ x^{n+1}D_{q}  \nonumber \\
J_{0} &=&-j+\frac{1}{n}\ xD_{q} \\
J_{-} &=&\frac{1}{n}\ x^{1-n}D_{q}  \nonumber
\end{eqnarray}

It is straightforward to prove that the new set of generators satisfy the
following commutations relations

\begin{eqnarray}
\left[ J_{0},\ J_{-}\right] &=&-\ J_{-}\ g(J_{0})  \nonumber \\
\left[ J_{0},\ J_{-}\right] &=&g(J_{0})\ J_{+} \\
\left[ J_{+},\ J_{-}\right] &=&f(J_{0})  \nonumber
\end{eqnarray}
where

\begin{equation}
g(J_{0})=j(1-q^{-n})+\frac{[n]}{n}q^{-n}+(1-q^{-n})\ J_{0}
\end{equation}

\begin{eqnarray}
f(J_{0}) &=&j^{2}(2-q^{n}-q^{-n})-\frac{j}{n}([n]+[-n]) \\
&&+(2j(1-q^{-n})+\frac{1}{n}([n]-[-n])\ J_{0}+(q^{n}-q^{-n})\ J_{0}^{2}
\nonumber
\end{eqnarray}
with

\[
\lbrack n]\equiv \frac{1-q^{n}}{1-q}
\]

This relations define the algebra $\mathcal{D}_{q}^{n}(su(2))$, which is a
special case of the general ones studied in [12]. Then the Casimir operator
is of the form
\begin{equation}
C_{q}=J_{-}J_{+}+h(J_{0})
\end{equation}
where $h(J_{0})$ is found to be

\begin{equation}
h(J_{0})=q^{n}\ J_{0}^{2}+\frac{[n]}{n}\ J_{0}
\end{equation}
After evaluating $J_{-}J_{+}$ one obtains for $C_{q}$

\begin{equation}
C_{q}=j\left( \frac{[n]}{n}+q^{n}j\right)
\end{equation}

Is easy to verify that the Delbecq-Quesne [12] condition for the existence
of the Casimir operator

\begin{equation}
h(J_{0})-h(J_{0}-g(J_{0}))=f(J_{0})
\end{equation}
is satisfied.

Our next goal is the algebraization of (31) following the steps of the
preceding section. We first perform a ``rotation'' of the form

\begin{equation}
\Psi _{q}(x)= \tilde{\Psi} _{q}(x)F(-a(x))
\end{equation}
where $F(-a(x))$ is such that

\begin{equation}
D_{q}F(-a(x))=-D_{q}(a(x))F(-a(x))
\end{equation}
in analogy with the undeformed case where $F$ was the exponential. In what
follows we will use the following notation

\begin{equation}
D_{q}f\equiv f^{^{\prime }},\quad f(qx)\equiv f^{\ast }
\end{equation}
Using the fact that $(fg)^{^{\prime }}=f^{^{\prime }}g+f^{\ast }g^{^{\prime
}}$ , equation (31) reads as follows

\begin{equation}
-\frac{1}{2}\widetilde{\Psi }^{^{\prime \prime }}+\frac{a^{^{\prime }}}{2}(%
\widetilde{\Psi }^{^{\prime }\ast }+\widetilde{\Psi }^{\ast ^{\prime }})+%
\frac{(a^{^{\prime \prime }}-a^{^{\prime }\ast }a^{^{\prime }})}{2}%
\widetilde{\Psi }^{\ast \ast }+V\widetilde{\Psi }=E\ \widetilde{\Psi }\
\end{equation}
where we have suppressed the index $\ q$ and the variable $x$ for
simplicity. Taking into account that $\widetilde{\Psi }^{^{\prime }}=(%
\widetilde{\Psi }-\widetilde{\Psi }^{\ast })/(1-q)x$ we get the following
relations

\begin{eqnarray*}
\widetilde{\Psi }^{\ast } &=&\widetilde{\Psi }-(1-q)\ x\ \widetilde{\Psi }%
^{^{\prime }} \\
\widetilde{\Psi }^{^{\prime }\ast } &=&\widetilde{\Psi }^{^{\prime }}-(1-q)\
x\ \widetilde{\Psi }^{^{\prime \prime }} \\
\widetilde{\Psi }^{\ast ^{\prime }} &=&q\widetilde{\Psi }^{^{\prime
}}-(1-q)\ q\ x\widetilde{\Psi }^{^{\prime \prime }}\  \\
\widetilde{\Psi }^{\ast \ast } &=&\widetilde{\Psi }-2(1-q)\ x\widetilde{\Psi
}^{^{\prime }}\ +(1-q)^{2}\ x^{2}\ \widetilde{\Psi }^{^{\prime \prime }}
\end{eqnarray*}
Making these changes we obtain from (44)

\begin{eqnarray}
H_{G}\widetilde{\Psi } &=&-\frac{1}{2}(1+a^{^{\prime }}(1-q^{2})\
x+(a^{^{\prime }\ast }a^{^{\prime }}-a^{^{\prime \prime }})(1-q^{2})\ x^{2})%
\widetilde{\Psi }^{^{\prime \prime }}  \nonumber \\
&&+\frac{1}{2}(a^{^{\prime }}(1+q)+2(a^{^{\prime }\ast }a^{^{\prime
}}-a^{^{\prime \prime }})(1-q)\ x)\widetilde{\Psi }^{^{\prime }}  \nonumber
\\
&&+(V-\frac{1}{2}(a^{^{\prime }\ast }a^{^{\prime }}-a^{^{\prime \prime }})\ )%
\widetilde{\Psi }  \nonumber \\
&=&E\widetilde{\Psi }
\end{eqnarray}
This is the expression of the rotated Hamiltonian that we will attempt to
equate to

\begin{equation}
H_{G}=\sum_{a,b=0,\pm ,a\geq b}C_{ab}\ J_{a}J_{b}+\sum_{a,b=0,\pm }C_{a}\
J_{a}
\end{equation}
Notice the difference with the non-deformed case, here the change of
variables $u\rightarrow x$ has been done from the beginning..

Equating the coefficients of $\widetilde{\Psi }^{^{\prime \prime }}$,$%
\widetilde{\Psi }^{^{\prime }}$and $\widetilde{\Psi }$ of (45) with those
that we get when acting with (46) on $\widetilde{\Psi }$, we get the
following relations

\begin{eqnarray}
&&C_{++}\ x^{2n+2}\ q^{n+1}-C_{+0}\ x^{n+2}\ q+(C_{00}-C_{+-}q^{-n})\ x^{2}\
q  \nonumber \\
&&+C_{0-}\ x^{2-n}\ q^{1-n}+C_{--}\ x^{2-2n}\ q^{1-n} \\
&=&-\frac{n^{2}}{2}(1+a^{^{\prime }}(1-q^{2})\ x+(a^{^{\prime }\ast
}a^{^{\prime }}-a^{^{\prime \prime }})\ (1-q^{2})\ x^{2})  \nonumber
\end{eqnarray}

\begin{eqnarray}
&&C_{++}(\frac{[n+1]}{n^{2}}-\frac{2j(1+q^{n})}{n})\ x^{2n+1}+(C_{+0}(\frac{%
3j}{n}-\frac{1}{n^{2}})-C_{+}\frac{1}{n})\ x^{n+1}  \nonumber \\
&&+(C_{+-}(\frac{2j}{n}-\frac{[1-n]}{n^{2}})+C_{00}(\frac{1}{n^{2}}-\frac{2j%
}{n})+C_{0}\frac{1}{n})\ x \\
&&+(C_{0-}(\frac{[1-n]}{n^{2}}-\frac{j}{n})+C_{-}\frac{1}{n})\ x^{1-n}+C_{--}%
\frac{[1-n]}{n^{2}}\ x^{1-2n}  \nonumber \\
&=&\frac{1}{2}(a^{^{\prime }}(1+q)+2(a^{^{\prime }\ast }a^{^{\prime
}}-a^{^{\prime \prime }})(1-q)\ x)  \nonumber
\end{eqnarray}

\begin{eqnarray}
&&C_{++}2j(2j-\frac{[n]}{n})\ x^{2n}+2j(C_{+}-C_{+0}j)\
x^{n}+j(C_{00}j-C_{0})  \nonumber \\
&=&V(x)-\frac{1}{2}(a^{^{\prime }\ast }a^{^{\prime }}-a^{^{\prime \prime }})%
\newline
\end{eqnarray}
>From the above relations we see that $a(x)$ can only be a combination of
powers in $x$, otherwise these expressions could not be satisfied.

Let us see the following case:\ $n=2,\ C_{++}=C_{--}=0.$ Furthermore we let $%
\ a^{^{\prime }}$ to be

\begin{equation}
a^{^{\prime }}=\omega x-\frac{[l+1]}{x}
\end{equation}
therefore : $a^{^{\prime \prime }}=\omega +\frac{[l+1]}{qx^{2}}$ and $\
a^{^{\prime }\ast }=\omega qx-\frac{[l+1]}{qx}.$ Then we obtain from (47)
the following conditions

\begin{eqnarray}
C_{+\,0} &=&2\omega ^{2}(1-q^{2})  \nonumber \\
C_{00}q^{2}-C_{+-} &=&2\omega \lbrack l+1](1-q^{4}) \\
C_{0-} &=&-2q-[l+1]\left( [l+1]-1-q\right) \left( 1-q^{2}\right)  \nonumber
\end{eqnarray}
>From (48) we have

\begin{eqnarray}
C_{+0}(3j-\frac{1}{2})-C_{+} &=&2\omega ^{2}q(1-q) \\
C_{+\,-}(2j+\frac{1}{2q})+C_{00}(\frac{1}{2}-2j)+C_{0} &=&2\omega \lbrack
l+1]\left( q+q^{-1}\right) \left( q-1\right)  \nonumber \\
&&+\omega \left( 3q-1\right)  \nonumber \\
C_{0-}(j+\frac{1}{2q})-C_{-} &=&[l+1]\left( 1+q\right)  \nonumber \\
&&+\frac{2}{q}[l+1]\left( [l+1]-1\right) \left( q-1\right)  \nonumber
\end{eqnarray}
and finally from (49)

\begin{eqnarray}
(C_{+}-C_{+0})2jx^{2}+(C_{00\ }j-C_{0})\ j &=&-\frac{1}{2}\omega ^{2}qx^{2}+%
\frac{[l+1](1-[l+1])}{2qx^{2}} \\
&&+\frac{1}{2}\omega ([l+1](q^{-1}+q)+1)+V(x)  \nonumber
\end{eqnarray}
In this way (51) and (52) provide us with six equations relating $C_{+0}$ , $%
C_{+}$, $C_{00\ }$, $C_{0}$, $C_{+\,-}$, $C_{0-}$ and \ $C_{-}$ \ with the
set $\left\{ \omega ,[l],q,j\right\} $. Since we have a set of six linear
equations and seven variables, we will be free to choose one of those
variables at will. Specifically, looking at the equations, we can see that $%
C_{+0}$, $C_{+}$, $C_{0-}$ and $C_{-}$ are fixed, and one of the variables $%
C_{00 }$, $C_{0}$ and $C_{+-}$ will be free, and fixing one of them we fix
the others two. This freedom on one of the variables corresponds in the
classical (non-deformed) case, to the freedom that we have in choosing the
ground state of the system.

>From (53) we get $V(x)$. We then need the coefficients $C_{+0}$ and $C_{+}$%
. The expression for $C_{+0}\ $\ is given by (51) and then from (52) we
obtain for $C_{+}$

\[
C_{+}=\omega ^{2}(1-q)(6j(1+q)-3q-1)
\]
then from (52) $V(x)$ is finally given by

\begin{equation}
V(x)=\omega _{q}^{2}x^{2}+\frac{1}{2q}\frac{[l+1]([l+1]-1)}{x^{2}}+V_{0}
\end{equation}
where $\omega _{q}^{2}$ and $V_{0}$ are defined as

\begin{eqnarray*}
\omega _{q}^{2} &=&\omega ^{2}\left( (1-q^{2})(8j^{2}-2j)-4jq(1-q)+\frac{q}{2%
} \right) \\
V_{0} &=&C_{00}j^{2}-C_{0}j-\frac{\omega }{2}-\frac{\omega \lbrack l+1]}{2}%
(q+q^{-1})
\end{eqnarray*}

We can say that $V(x)$ correspond to a q-deformed harmonic oscillator with a
q-deformed angular frequency $\omega _{q}$ and with a q-deformed angular
momentum. Notice that in the limit $q\rightarrow 1$ we obtain the classical
potential.

%{\Large Deformed spectrum and eigenfunctions}

\section{Deformed spectrum and eigenfunctions}

\smallskip $\smallskip$

As we mentioned before, the carrier space of $SU(2)$, i.e. $R^{j}$, is given
in the classical case by monomials on the variable $u$. This was the
original variable defining the algebra, see (7) and (8). We can use this
basis in the deformed case. Since we have chosen a change of variables given
by $u=x^{2}$, the monomials will be of the form $x^{2k}$ ($k$ integer),
these are represented as $\left| k\right\rangle $. Then from (33) we have

\begin{eqnarray}
J_{+}\left| k\right\rangle &=& \left( 2j-\frac{[2k]}{2}\right) \left|
k+1\right\rangle \\
J_{0}\left| k\right\rangle&=& \left( -j-\frac{[2k]}{2}\right) \left|
k\right\rangle \\
J_{-}\left| k\right\rangle&=&\frac{[2k]}{2}\left| k-1\right\rangle
\end{eqnarray}

If $\ j=\frac{1}{4}[2m]$ for some integer $m$, the representation will be
finite-dimensional, otherwise will be infinite-dimensional and bounded
below. Since we have expressed $H_{G}$ in terms of the generators of the
algebra, then we can represent it as a matrix acting on the corresponding
basis. If the representation is infinite-dimensional it will be too hard to
diagonalize $H_{G}$ in order to obtain the spectrum, therefore, we will work
with the finite case, so we have only to diagonalize a finite matrix as
happened in the classical case.

According to (41), the eigenfunctions will be of the form,

\[
\Psi (x)=P(x^{2})F(-a(x))
\]
where $P(x^{2})$ is a polynomial on $x^{2}$.Our next task is to evaluate the
function $F(-a(x)).$ Let us write $F(-a(x))=x^{l+1}f(x)$, then from
equations (42) and (50) we get

\[
D_{q}(x^{l+1}f(x))=\left( -\omega x+\frac{[l+1]}{x}\right) f(x)
\]
we obtain
\begin{equation}
q^{l+1}x^{l+1}D_{q}(f(x))=-\omega x^{l+2}f(x)
\end{equation}
if we write $f(x)$ as a power series $f(x)=\stackrel{\infty }{%
\sum\limits_{n=0}}\alpha _{n}x^{n}$, the coefficients $\alpha _{n}$ are
found to satisfy the following conditions

\begin{eqnarray*}
\alpha _{n} &=&-\frac{\omega }{q^{l+1}[n]}\alpha _{n-2},\quad
\mbox{for {\it
n} even} \\
&=&0,\quad \mbox{for {\it n} odd}
\end{eqnarray*}
Hence

\[
\alpha _{2n}=\alpha _{0}\left( -\frac{\omega (1+q)}{q^{l+1}}\right) ^{n}%
\frac{1}{[n]_{q^{2}}!}
\]
where $\alpha _{0}$\ is arbitrary and $[n]_{q^{2}}$ is defined in the same
way as $[n]$, but with the replacement $q\rightarrow q^{2}$, and the
factorial means

\[
\lbrack n]_{q^{2}}!=[n]_{q^{2}}[n-1]_{q^{2}}\cdots
\]
With these results we see that the function\ $f(x)$ is
\[
f(x)=\sum_{n=0}^\infty \frac{1}{[n]_{q^{2}}!}\left( -\frac{\omega (1+q)}{%
q^{l+1}}x^{2}\right)^n
\]
which is a deformed exponential with deformation parameter $q^{2}$ and with
argument $-\frac{\omega (1+q)}{q^{l+1}}x^{2}$. So the function $F(-a(x))$
turns out to be

\begin{equation}
F(-a(x))=x^{l+1}\exp _{q^{2}}\left( -\frac{\omega (1+q)}{q^{l+1}}x^{2}\right)
\end{equation}
where we have set $\alpha _{0}=1$.

In order to evaluate the energy spectrum we need the matrix elements of the
deformed Hamiltonian. These elements are easy to compute from the
expressions (46) together with (55), (56) and (57), we get

\begin{eqnarray}
\left\langle n\left| H_{G}\right| k\right\rangle &=&\left( 2j-\frac{[2k]}{2}%
\right) \left( C_{+}+C_{+0}\left( (-j+\frac{[2k]}{2})\right) \right) \delta
_{n,k+1} \\
&&+\left( C_{+-}\left( \frac{[2k]}{2}\right) \left( 2j-\frac{[2k]}{2}\right)
+C_{00}\left( -j+\frac{[2k]}{2}\right) ^{2}+C_{0}\left( -j+\frac{[2k]}{2}%
\right) \right) \delta _{n,k}  \nonumber \\
&&+\left( \frac{[2k]}{2}\right) \left( C_{0-}\left( -j+\frac{[2k]}{2}\right)
+C_{-}\right) \delta _{n,k-1}  \nonumber
\end{eqnarray}

To illustrate the method, let's see some examples. The simplest possible
cases are $j=0$ and $j=\frac{1}{4}[2]$. In the case $j=0$\ we have obviously
one energy eigenvalue, which is $E=0$ and its corresponding eigenfunction is
then

\[
\Psi _{0}(x)= x^{l+1}\exp _{q^{2}}\left( -\frac{\omega (1+q)}{q^{l+1}}%
x^{2}\right)
\]
For\ $j=\frac{1}{4}[2]=\frac{1}{4}(1+q)$, the representation space will be
of dimension 2 and will be given by $R^{j}=\{1,x^{2}\}$. As it was said
before, we still have a free parameter ($C_{+-}$, $C_{00}$ or $C_0$). For
convenience, we choose the fixing condition given by
\[
C_{00}j-C_{0}=0
\]
In this way, the matrix representation of $H_{G1}$ is nicely given by
\begin{equation}
H_{G1}=\left(
\begin{array}{cc}
0 & 2j(C_{-}-jC_{0-}) \\
2j(C_{+}-jC_{+0}) & 2j(C_{0}+2jC_{+-})
\end{array}
\right)
\end{equation}

So obtaining the values for $C_{+0}$, $C_{+-}$, $C_{0-}$, $C_{+}$, $C_{0}$
and $C_{-}$ from (51) and (52), substituting them in the above matrix, and
diagonalizing it, we can get two energy eigenvalues for the system in
question, as well as the corresponding eigenfunctions. The two eigenvalues
are the following ones:

\[
E_{q\pm }=\left( p_{1}[l+1]+p_{2}\pm 2\sqrt{p_{3}[l+1]^{2}+p_{4}[l+1]+p_{5}}%
\right) \frac{(1+q)}{8q}\omega
\]
where $p_{i}$, $i=1\ldots 5$, are polynomials on $q$ given by

\begin{eqnarray}
p_1 & = & -2(q^2+1)(q-1)^2  \nonumber \\
p_2 & = & 2q(3q-1)  \nonumber \\
p_3 & = & (1+q)^2(q-1)^6  \nonumber \\
p_4 & = & 2q(-1+2q)(1+q)^2(q-1)^3  \nonumber \\
p_5 & = & q^2(1+q)(4q^2-3q+1)  \nonumber
\end{eqnarray}

We can see that in the limit $q \rightarrow 1$ we get the expected
eigenvalues for the resulting potential:

\[
V(x) \rightarrow \frac{1}{2}\omega^2 x^2 + \frac{l(l+1)}{2x^2}+ \left( \frac{%
3}{2}- l \right) \omega
\]
i.e.

\begin{eqnarray*}
E_{q-}& \rightarrow& 0 \\
E_{q+}& \rightarrow& 2\omega
\end{eqnarray*}

It is possible also to get the eigenfunctions associated to each eigenvalue.
They will be of the form

\[
\Psi _{\pm}(x)=\left( A_{\pm}+B_{\pm}x^2 \right) x^{l+1}\exp _{q^{2}}\left( -%
\frac{\omega (1+q)}{q^{l+1}}x^{2}\right)
\]
where $A_{\pm}$ and $B_{\pm}$ are some complicated functions of $\omega$ and
$[l+1]$, which we will not show here.

\section{Final Comments}

We have seen how using a q-algebra of the kind studied in [12] we can get
exact solutions for part of the energy eigenvalues and eigenfunctions of the
q-Schr\"{o}dinger equation in the case in which the potential is a
3-dimensional q-deformed harmonic oscillator. The deformation of the
potential is a very specific one, i.e. with a very specific q-dependence, as
shown in (54). We could also want to construct any q-dependent potential
such that in the limit $q \rightarrow 1$ we get the usual classical
potential. Our solution is only valid for the specific case we was
considering, so a general solution for any possible choice of the
q-dependence remains as an open and very hard question. Even harder is to
consider potentials others than a harmonic oscillator, like the Morse
potential or so. This is because of the somehow complicated properties of
the deformed derivative.

A remarkable aspect of our solution is the fact that we can get only part of
the spectrum, the rest remaining unknown. This shouldn't be too astonishing,
after all our method is based on the ideas of partial algebraization, which
in most of the cases allows us to get only part of the spectrum, the only
exceptions being the ones studied at the end of section 2. Nevertheless,
considering that one of those exceptions was precisely the harmonic
oscillator, it is curious that we cannot get the whole spectrum in the
deformed case, but only a part of it.


\begin{thebibliography}{99}
\bibitem{}  Natanzon G A 1979 \textit{Theor. Mat. Fiz.} \textbf{38} 146

\bibitem{key2}  Wu J, Alhassid Y and F G\"{u}rsey 1989 \textit{Ann. Phys.}
\textbf{196} 163; Wu J and Alhassid Y 1990 \textit{J. Math. Phys.} \textbf{31%
} 557

\bibitem{key3}  Gendenshtein L E 1983 \textit{JEPT Lett.} \textbf{38} 356;
Witten E 1981 \textit{Nucl. Phys.} \textbf{B 185} 513; Salomonson P and van
Holten W 1982 \textit{Nucl. Phys.} \textbf{B 196} 509; Cooper F, Ginocchio J
N and Khare A 1987 \textit{Phys. Rev. D} \textbf{36 } 2458. A rather
complete rewiew can be found in Cooper F, Khare A and Sukhatme U 1995
\textit{Phys. Rep}. \textbf{251 }267

\bibitem{key4}  Cordero P and Salam\'{o} S 1993 \textit{Found. Phys.}
\textbf{23 } 675; 1994 \textit{J. Math. Phys.} \textbf{35} 3301; 1991
\textit{J. Phys. A: Math.Gen.} \textbf{24} 5299

\bibitem{key5}  Turbiner A V and Ushveridze A\ G 1987 \textit{Phys. Lett.}
\textbf{126A} 181; Turbiner A V 1988 \textit{Comm. Math. Phys.} \textbf{118}
467; Shifman M\ A 1989 \textit{Int. J. Mod. Phys.} \textbf{A 4} 2897

\bibitem{key6}  Kosteleck\'{y} V A, Man'ko V\ I, Nieto M M and Truax D R
1993 Phys. Rev. A \textbf{48} 951

\bibitem{key7}  Dayi \"{O} F and Duru I H 1997 Tr. J. of Phys. \textbf{21}
348

\bibitem{key8}  Cooper I L and Gupta R K 1995 \textit{Phys. Rev. A} \textbf{%
52} 941

\bibitem{key9}  Daskaloyannis C J 1992 \textit{J. Phys. A: Math. Gen.}
\textbf{25} 2261; Bonatsos D and Daskaloyannis C 1992 \textit{Phys. Rev. A}
\textbf{46} 76; Bonatsos D, Daskaloyannis C, Kolokotronis P and Lenis D,
1997 \textit{Tr. J. of Physics} \textbf{21} 303

\bibitem{key10}  De Freitas A and Salam\'{o} S, 1999 \textit{\ Nuovo Cimento
B }\textbf{114} 535

\bibitem{ke11}  Dabrowska J W, Khare A and Sukhatme U P 1988 \textit{J.
Phys. A: Math. Gen} \textbf{21} L 195

\bibitem{key12}  Delbecq C and Quesne C 1993 \textit{J. Phys. A: Math. Gen.}
\textbf{26} L 127; 1993 \textit{Phys. Lett.} \textbf{B 300 } 227
\end{thebibliography}
\end{document}